\documentclass[a4paper]{article}
\usepackage[T1]{fontenc}
\usepackage[latin9]{inputenc}
\usepackage{amsmath}
\usepackage{amsthm}
\usepackage{amssymb}
\usepackage{esint}

\makeatletter

\usepackage[all]{xy}
\usepackage{graphicx}

\makeatother

\begin{document}
\begin{flushleft}
\textbf{OUJ-FTC-9}\\
\par\end{flushleft}

\begin{center}
{\LARGE{}{}Quantization of Nambu Brackets from Operator Formalism
in Classical Mechanics}{\LARGE\par}
\par\end{center}

\begin{center}
 
\par\end{center}

\begin{center}
\vspace{32pt}
 So Katagiri\textsuperscript{}\footnote{So.Katagiri@gmail.com} 
\par\end{center}

\begin{center}
\textit{Nature and Environment, Faculty of Liberal Arts, The Open
University of Japan, Chiba 261-8586, Japan} 
\par\end{center}

\vspace{10pt}

This paper proposes a novel approach to quantizing Nambu brackets
in classical mechanics using operator formalism. The approach employs
the \textquotedblleft Planck derivative\textquotedblright{} to represent
Nambu brackets, from which we derive a commutation relation for their
quantization. Notably, this commutation relation aligns with that
emerging from the T-duality of closed strings in a twisted torus with
a B-field, thereby hinting at a potential connection with Double Field
Theory. 

\section{Introduction}

Clarifying the relationship between classical and quantum theory is
an activity in which the understanding of the macroscopic world from
the perspective of the microscopic world and the understanding of
the microscopic world from the perspective of the macroscopic world
are mutually reconstructed. It is an actual problem for the foundation
of modern physics, a straightforward example of which is the quantization
of Poisson brackets. The quantization of more complex structures,
such as Nambu brackets, is still controversial and continues to be
an exciting and challenging problem that ties together more profound
questions about perspectives on the world in theoretical physics.

The Nambu bracket, introduced by Yoichiro Nambu in 1973\cite{Nanbu},
is an extension of the Poisson bracket to higher dimensions and later
applied in various fields, including integrable systems, fluid dynamics,
nonequilibrium thermodynamics, and M-theory, and is particularly important
in describing membranes\cite{Sugamoto,NanbuFluid,SugamotoSaito,KatagiriSugamoto}.

We can point out three main features inherent to the Nambu bracket: 
\begin{itemize}
\item It is based on multiple Hamiltonians representing conserved quantities. 
\item It follows the Leibniz rule of the time derivative. 
\item It satisfies a generalized Jacobi identity called the fundamental
identity (F.I.). 
\end{itemize}
Quantization of Nambu brackets, however, is difficult, and most previous
attempts have violated at least one of its inherent properties of
Nambu brackets \cite{AwataYoneda,MinicDeformation,CurtrightZachos,MinicHsiung,TakhtajanFlato,MatsuoHo-2,key-25},
see \cite{YoneyaReview} for a review.

Operator formalism in classical mechanics describes classical mechanics
in Hilbert spaces studied by Koopman\cite{koopman} and von Neumann\cite{neumann}
in 1931, a formalism that is now much neglected in physics research.
Recently, an attempt has been made to describe quantum mechanics in
such a formalism\cite{katagiri,Bondar}. Such attempts have advanced
our understanding of the quantization process by viewing quantization
not as a conversion from c-numbers to q-numbers but rather as a two-step
conversion process to operator formalism and introducing quantum corrections.

This research explores a new method of quantizing Nambu brackets using
operator formalism in classical mechanics.

Central to our method is introducing of the Planck derivative (difference)
$D$, which represents the difference between the quantum and classical
operators when divided by the Planck constant. From this representation
of Nambu brackets by $D$, we derive the commutation relation that
must be satisfied by the quantization of Nambu brackets. Coincidentally,
the same commutation relation appears in the discussion of noncommutativity
obtained by the T dual of a closed string in a twisted torus with
a B field, suggesting a connection with Double Field Theory.

The paper is organized as follows: Section 2 gives an overview of
Nambu brackets and their role in M-theory. Section 3 introduces the
classical operator formalism and explains its importance in the context
of our research. Section 4 introduces the Planck derivative to link
the operator formalism of classical mechanics with quantum mechanics.
Section 5 describes our proposed approach to quantizing of Nambu brackets
using classical operator formalism and discusses the resulting commutation
relation. Section 6 applies the quantization of Nambu brackets to
several examples. Finally, in Section 7, we combine our conclusions
with an analysis of the study's results and point out possible directions
for future research.

Appendix A also provides details of the commutation relation. 

This research is not only a contribution to the quantization of the
M2-brane, a pivotal component of M-theory\cite{IKKT,BFSS,BLG,BLG2},
but also has the potential to stimulate further consideration of the
nature of quantum theory from a physical and fundamental point of
view: namely, what quantization is.

\section{Nambu bracket and M-theory}

In Hamiltonian mechanics, the state space is spanned by two variables,
$x$ and $p$. Nambu mechanics is an extension of this state space
to $n$ variables, denoted as $x^{i}$. In this context, we consider
the case of $n=3$.

The time evolution of mechanical variables is described using multiple
``Hamiltonian'' as 
\begin{equation}
\dot{x}^{i}=\{H_{1},\ H_{2},\ x^{i}\},
\end{equation}
where $\{A,B,C\}$ denotes the Nambu bracket, defined as:

\begin{equation}
\{A,B,C\}\equiv\epsilon^{ijk}\frac{\partial A}{\partial x^{i}}\frac{\partial B}{\partial x^{j}}\frac{\partial C}{\partial x^{k}}.
\end{equation}
Nambu mechanics possesses three main features\cite{MatsuoHo-2}. First,
the relation

\begin{equation}
\{A_{\sigma(1)},A_{\sigma(2),}A_{\sigma(3)}\}=(-1)^{\sigma}\{A_{1},A_{2},A_{3}\},
\end{equation}
is based on multiple Hamiltonians that represent conserved quantities
of the system. These Hamiltonians are employed to describe the system's
dynamics in terms of its state variables.

Second, 
\begin{equation}
\{AB,C,D\}=A\{B,C,D\}+\{A,C,D\}B,
\end{equation}
which signifies that Nambu mechanics adheres to the Leibniz rule for
time derivatives. This rule determines how the system's variables
change over time. This rule ensures that the time evolution of the
system is consistent with the underlying physical laws.

Finally, 
\begin{equation}
\{\{A,B,C\},D,E\}=\{\{A,D,E\},B,C\}+\{A,\{B,D,E\},C\}+\{A,B,\{C,D,E\}\},
\end{equation}
denoted as the fundamental identity (F.I.), signifies that Nambu mechanics
satisfies a generalized Jacobi identity. This identity guarantees
the consistency and well-defined nature of the Nambu bracket operation,
which is crucial for describing the behavior of complex physical systems.

Membranes serve as fundamental objects in the description of M-theory.
The Nambu bracket is closely connected to membrane theory in its Lagrangian
formalism\cite{Sugamoto,key-25}, as elaborated below.

In the mechanics of a point particle, the action can be expressed
as follows:

\begin{equation}
\begin{aligned}S & =\int_{C_{1}}pdx-Hdt=\int_{C_{2}}dp\wedge dx-dH\wedge dt\\
 & =\int_{C_{2}}\left(\epsilon_{ij}\epsilon^{ab}\partial_{a}x^{i}\partial_{b}x^{j}-\frac{\partial H}{\partial x^{i}}\partial_{\sigma}x^{i}\right)d\sigma dt\\
 & =\int_{C_{2}}\left(\epsilon_{ij}\{x^{i},x^{j}\}-\{H,t\}\right)d\sigma dt,
\end{aligned}
\end{equation}
where $C_{1}=\partial C_{2},$ $x^{1}=x,\ x^{2}=p$, and the subscripts
$i$ are $i=x,p$ and $a$ are $\ a=t,\sigma$. The symbol $\epsilon_{ij}\ $
and $\epsilon_{ab}$ represent the Levi-Civita symbol.

The dynamics of a point particle can be represented in two dimensions,
where the additional parameter can be interpreted as a parameter of
change arising from virtual work. Consequently, in this context, the
dynamics of a point particle possess an inherent description akin
to closed string theory, characterized by two parameters.

In Nambu dynamics, the action is expressed as:
\begin{equation}
S=\int_{C_{3}}\left(\epsilon_{ijk}\epsilon^{abc}\partial_{a}x^{i}\partial_{b}x^{j}\partial_{c}x^{k}-\frac{\partial H_{1}}{\partial x^{i}}\frac{\partial H_{2}}{\partial x^{j}}\partial_{[\sigma_{1}}x^{i}\partial_{\sigma_{2}]}x^{j}\right)d^{2}\sigma dt.
\end{equation}

The original action originally included two string-like parameters,
but it can be rewritten as a three-parameter membrane-like action
by introducing an additional virtual displacement parameter.\footnote{In particular, we can determine the form of the Hamiltonian so that
it satisfies the following,

\begin{equation}
\{H,t\}\equiv e(\epsilon_{ij}\epsilon^{ij}+1)
\end{equation}
where $e$ is the Lagrange multiplier and if $\epsilon_{ij}$ depends
on $(t,\sigma)$, the action is 
\begin{equation}
S_{s}=\int_{C_{2}}\left(\epsilon_{ij}\{x^{i},x^{j}\}-e\epsilon_{ij}\epsilon^{ij}-e\right)d\sigma dt.
\end{equation}

We integrate out $\epsilon_{ij}$ and obtain 
\begin{equation}
S_{s}=\int_{C_{2}}\left(\frac{1}{e}\{x^{i},x^{j}\}\{x_{i},x_{j}\}-e\right)d\sigma dt.
\end{equation}

In Nambu dynamics, we rewrite as 
\begin{equation}
S=\left(\int_{C_{3}}\epsilon_{ijk}\{x^{i},x^{j},x^{k}\}-\{H_{1},H_{2},t\}\right)d^{2}\sigma dt.
\end{equation}
If we set the form of the Hamiltonian so that it satisfies the following,
\begin{equation}
\{H_{1},H_{2},t\}\equiv e\left(\epsilon_{ijk}\epsilon^{ijk}+1\right)\label{eq:membraneH1H2}
\end{equation}
where $e$ is the Lagrange multiplier and if $\epsilon_{ijk}$ depends
on $(t,\sigma_{1},\sigma_{2})$, the action is 
\begin{equation}
S_{m}=\int_{C_{3}}\left(\epsilon_{ijk}\{x^{i},x^{j},x^{k}\}-e\epsilon_{ijk}\epsilon^{ijk}-e\right)d^{3}\sigma.
\end{equation}

We integrate out $\epsilon_{ijk}$ and obtain 
\begin{equation}
S_{m}=\int_{C_{3}}\left(\frac{1}{e}\{x^{i},x^{j},x^{k}\}\{x_{i},x_{j},x_{k}\}-e\right)d^{3}\sigma,
\end{equation}
which is the relativistic membrane action.}

From the above, it can be observed that Nambu dynamics is associated
with a $(D-1)$-dimensional membrane where $D$ represents the number
of dynamical variables.

\section{Operator formalism in classical mechanics}

The operator formalism of classical mechanics was initially developed
in the 1930s by Koopman\cite{koopman} and von Neumann\cite{neumann}.
However it has undergone revisions in recent years and offers the
advantage of dealing with hybrid systems that combine classical and
quantum theory\cite{katagiri,Bondar,mauro,gozzi,Manjarres,Theuel}.
A key characteristic of the operator formalism of classical mechanics
is the differentiation between quantum and classical operators.

In quantum mechanics, the relationship between position and momentum
in Poisson brackets

\begin{equation}
\{x^{i},x^{j}\}=\epsilon^{ij}
\end{equation}
is quantized by replacing it with the corresponding equation 
\begin{equation}
[\hat{x}^{(q)i},\hat{x}^{(q)j}]=i\hbar\epsilon^{ij}.
\end{equation}
Here, the variable is denoted with a superscript $(q)$ to emphasize
that it represents a quantum physical quantity.

The operator formalism of classical mechanics is characterized by
replacing position and momentum with operators while preserving their
commutativity

\begin{equation}
[\hat{x}^{(c)i},\hat{x}^{(c)j}]=0,
\end{equation}
and introducing their corresponding canonical counterparts, denoted
as $\hat{\xi}_{i}$:

\begin{equation}
[\hat{x}^{(c)i},\hat{\xi}_{j}]=i\delta_{j}^{i},\ [\hat{\xi}_{i},\hat{\xi}_{j}]=0.
\end{equation}
Additionally, the Liouvillian operator, denoted as $\hat{L}$, is
introduced in correspondence with the Hamiltonian:

\begin{equation}
\hat{L}=\epsilon^{ij}\frac{\partial H(\hat{x})}{\partial\hat{x}^{i}}\hat{\xi}_{j}.
\end{equation}

Since position and momentum are commute and can be simultaneously
diagonalized by eigenstates, the wave function can be expressed as
a function of position and momentum:
\begin{equation}
|\psi\rangle=\int dxdp|x,p\rangle\langle x,p|\psi\rangle.
\end{equation}
The Schr$\mathrm{\ddot{o}}$dinger equation can be written as:

\begin{equation}
i\frac{\partial}{\partial t}|\psi\rangle=\hat{L}|\psi\rangle.
\end{equation}
As a specific example, for a free particle\cite{mauro}, the Liouvillian
operator $\hat{L}$ can be represented as 
\begin{equation}
\hat{L}=\frac{\hat{p}}{m}\hat{\xi}_{x}.
\end{equation}
The corresponding Schr$\mathrm{\ddot{o}}$dinger equation becomes:
\begin{equation}
i\frac{\partial}{\partial t}|\psi\rangle=\int d\xi'_{x}dp'\frac{p'}{m}\xi'_{x}|\xi'_{x},p'\rangle\langle\xi'_{x},p'|\psi\rangle.
\end{equation}
By applying $\langle\xi_{x},p|$ from left-hand side, we obtain, 
\begin{equation}
i\frac{\partial}{\partial t}\langle\xi_{x},p|\psi\rangle=\frac{p}{m}\xi_{x}\langle\xi_{x},p|\psi\rangle.
\end{equation}
By introducing a proportionality factor $A$, the solution can be
expressed as:

\begin{equation}
\langle\xi_{x},p|\psi\rangle=Ae^{i\frac{p}{m}\xi_{x}t}.
\end{equation}
Performing the Fourier transform with respect to $\xi_{x}$, we obtain:
\begin{equation}
\langle x,p|\psi\rangle=A\delta(x-\frac{p}{m}t).
\end{equation}
This solution reproduces the linear trajectory of a free particle
in classical mechanics.

\section{Planck derivative $D$}

In the previous section, we discussed the operator formalism of classical
mechanics, where we observed the commutativity of position and momentum
operators. However, the Poisson bracket-like structures that are essential
in canonical quantization are not explicitly present in this framework.
To establish a connection between the operator formalism of classical
mechanics and quantum mechanics, it is important to consider structures
analogous to Poisson brackets within the classical operator formalism.

To establish the relationship between the operator formalism of classical
mechanics and quantum mechanics, we introduce the difference between
classical and quantum operators denoted as $D$:

\begin{equation}
\hat{x}^{(q)i}=\hat{x}^{(c)i}+\hbar D(\hat{x}^{(c)i}),
\end{equation}

We refer to $D$ as the Planck derivative,

\begin{equation}
D=\frac{\hat{x}^{(q)i}-\hat{x}^{(c)i}}{\hbar}.
\end{equation}

Next, we can express the Poisson bracket within the operator formalism
of classical mechanics using $D$ as follows:\footnote{We can also check that the commutation relation x$^{q}$ satisfies

\[
[\hat{x}^{(q)i},\hat{x}^{(q)j}]=\frac{1}{2}i[\hat{x}^{(c)i},-2i\hbar D(\hat{x}^{(c)})^{j}]+\frac{1}{2}i[-2i\hbar D(\hat{x}^{(c)})^{i},\hat{x}^{(c)j}]
\]

\[
=\frac{1}{2}i\hbar\epsilon^{ij}+\frac{1}{2}i\hbar\epsilon^{ij}=i\hbar\epsilon^{ij}.
\]
}

\begin{equation}
[\hat{x}^{(c)i},-2iD(\hat{x}^{(c)j})]=\epsilon_{ij}.
\end{equation}

Note that the right-hand side of the equation represents the usual
antisymmetric epsilon symbol, indicating the Poisson bracket relationship
between the position operator $\hat{x}^{(c)i}$ and the Planck derivative
$-2iD(\hat{x}^{(c)j})$.

The results of the previous section indicate that $D$ can be expressed
as:

\begin{equation}
D(\hat{x}^{(c)j})=-\frac{1}{2}\epsilon^{jk}\hat{\xi}_{k}.
\end{equation}

Therefore, the Schr$\mathrm{\ddot{o}}$dinger equation of quantum
mechanics can be alternatively expressed using classical operators\cite{katagiri}:

\begin{equation}
i\hbar\frac{\partial}{\partial t}|\psi^{(q)}\rangle=H\left(\hat{x}^{(c)i}-\frac{1}{2}\hbar\epsilon^{ij}\hat{\xi}_{j}\right)|\psi^{(q)}\rangle.
\end{equation}

In this equation,$H$ represents the Hamiltonian, and$|\psi^{(q)}\rangle$
denotes the quantum state. The expression inside the parentheses indicates
the modified classical operators incorporating the Planck derivative
term.

Note that there exists a well-known approach to the representation
of operators in phase space called the Bopp operator\cite{key-5,key-4,key-9,key-3}.
Also, the Schr$\mathrm{\ddot{o}}$dinger equation on such a phase
space has already been examined and discussed in previous work \cite{key-6,key-7}.

\subsection{Dual operator of $\hat{x}^{(q)i}$}

In this section, we will delve into the concept of the dual operator
associated with $\hat{x}^{(q)i}$. The dual operator, denoted as $\hat{\tilde{x}}^{(q)i}$,
is obtained by interchanging the classical and the Planck derivative
operator $D(\hat{x}^{(c)i})$:

\begin{equation}
\hat{\tilde{x}}^{(q)i}=\hat{x}^{(c)i}-\hbar D\left(\hat{x}^{(c)i}\right),
\end{equation}

\begin{equation}
D(\hat{x}^{(c)i})=\frac{\hat{x}^{(c)i}-\hat{\tilde{x}}^{(q)i}}{\hbar}.
\end{equation}
Notably, these operators independently satisfy the Heisenberg algebra,
reflecting the fundamental commutation relations of quantum mechanics:\footnote{
\[
[\hat{x}^{(c)i}-\hbar D(\hat{x}^{(c)i}),\ \hat{x}^{(c)j}-\hbar D(\hat{x}^{(c)j})]=-\frac{1}{2}i[\hat{x}^{(c)i},-2i\hbar D(\hat{x}^{(c)j}]+-\frac{1}{2}i[-2i\hbar D(\hat{x}^{(c)i}),\hat{x}^{(c)j}]
\]

\[
=-\frac{1}{2}i\hbar\epsilon^{ij}+\frac{1}{2}i\hbar\epsilon^{ji}=-i\hbar\epsilon^{ij}
\]
}

\begin{equation}
[\hat{\tilde{x}}^{(q)i},\hat{\tilde{x}}^{(q)j}]=-i\hbar\epsilon^{ij}
\end{equation}
but are commute with $\hat{x}^{(q)}$,\footnote{
\[
[\hat{x}^{(c)i}-\hbar D(\hat{x}^{(c)i}),\ \hat{x}^{(c)j}+\hbar D(\hat{x}^{(c)j})]=-\frac{1}{2}i[\hat{x}^{(c)i},+2i\hbar D(\hat{x}^{(c)j}]+-\frac{1}{2}i[-2i\hbar D(\hat{x}^{(c)i}),\hat{x}^{(c)j}]
\]

\[
=+\frac{1}{2}i\hbar\epsilon^{ij}+\frac{1}{2}i\hbar\epsilon^{ji}=0
\]
}

\begin{equation}
[\hat{\tilde{x}}^{(q)i},\hat{x}^{(q)j}]=0.
\end{equation}

See Appendix A for detailed calculations.

Consequently, the classical operator $\hat{x}^{(c)i}$ can be regard
as the center of mass between the quantum operator $\hat{x}^{(q)i}$
and the dual operator $\hat{\tilde{x}}^{(q)i}$:

\begin{equation}
\hat{x}^{(c)i}=\frac{\hat{x}^{(q)i}+\hat{\tilde{x}}^{(q)i}}{2}.
\end{equation}
Furthermore, the Planck derivative $\hbar D(\hat{x}^{(c)i})$ represents
their relative coordinate:
\begin{equation}
\hbar D(\hat{x}^{(c)i})=\frac{\hat{x}^{(q)i}-\hat{\tilde{x}}^{(q)i}}{2}.
\end{equation}

In summary, in classical mechanics, states are represented by simultaneous
eigenstates of position and momentum,

\begin{equation}
|x^{(c)},p^{(c)}\rangle
\end{equation}
or 
\begin{equation}
|x^{(c)},\xi_{p}\rangle,\ \mathrm{or}\ |\xi_{x},p^{(c)}\rangle\ \mathrm{or}\ |\xi_{x},\xi_{p}\rangle.
\end{equation}
On the other hand, in the equivalent quantum mechanics, states are
described by simultaneous eigenstates of posision $x$ and its dual
coordinate $\tilde{x}$:

\begin{equation}
|x^{(q)},\tilde{x}^{(q)}\rangle
\end{equation}
or 
\begin{equation}
|p^{(q)},\tilde{p}^{(q)}\rangle,\ \mathrm{or}\ |x^{(q)},\tilde{p}^{(q)}\rangle\ \mathrm{or}\ |p^{(q)},\tilde{x}^{(q)}\rangle.
\end{equation}
As a result, the wavefunction in quantum mechanics includes both coordinates
and dual coordinates:

\begin{equation}
\psi^{(q)}(x^{(q)},\tilde{x}^{(q)}).
\end{equation}
By recombining these coordinates, similar to the Wigner function,
we obtain:

\begin{equation}
\psi^{(q)}\left(\frac{x^{(q)}+\tilde{x}^{(q)}}{2},x^{(q)}-\tilde{x}^{(q)}\right)=\psi^{(q)}(x^{(c)},2\hbar D(x))=\psi^{(q)}(x^{(c)},\hbar\xi_{p}),
\end{equation}
which can be viewed as the wavefunction of classical mechanics. Moreover,
by performing a Fourier transform on the second term, we obtain:

\begin{equation}
\psi^{(q)}(x^{(c)},p^{(c)}),
\end{equation}
representing the wavefunction of classical mechanics.

In conclusion, by incorporating dual coordinates, quantum mechanics
becomes entirely equivalent to classical mechanics.

\section{Quantization of Nambu brackets}

This section presents the main result of the paper.

In the operator formalism of classical mechanics, we introduced the
operator $D$ to preserve the Poisson brackets and establish a relationship
between quantum mechanical operators and classical mechanical operators.
In a similar manner, we introduce $D$ to preserve the Nambu bracket
in Nambu mechanics:

\begin{equation}
[\hat{x}^{(c)i},[\hat{x}^{(c)j},-2D(\hat{x}^{(c)k})]]=\epsilon^{ijk}.
\end{equation}
To satisfy this relationship, we require the following condition:

\begin{equation}
[\hat{x}^{(c)j},D(\hat{x}^{(c)k})]=\frac{1}{2}\epsilon^{ijk}\xi_{i}.
\end{equation}
Similar to classical mechanics, we define the quantum operator as:

\begin{equation}
\hat{x}^{(q)i}=\hat{x}^{(c)i}+\hbar D\left(\hat{x}^{(c)i}\right),
\end{equation}

\begin{equation}
D=\frac{\hat{x}^{(q)i}-\hat{x}^{(c)i}}{\hbar}.
\end{equation}
From the given definitions, we can derive the following commutation
relations for quantum operators:

\begin{equation}
[\hat{x}^{(q)i},\hat{x}^{(q)j}]=i\hbar\epsilon^{ijk}\hat{\xi}_{k},
\end{equation}

\begin{equation}
[\hat{x}^{(q)i},[\hat{x}^{(q)j},\hat{x}^{(q)k}]]=i\hbar\epsilon^{ijk}.
\end{equation}

Similar to ordinary quantum mechanics, the time evolution of operators
in the Heisenberg representation can be described as:

\begin{equation}
i\frac{\partial}{\partial t}\hat{O}^{(q)}=\frac{1}{\hbar}[\hat{O}^{(q)},[H_{1}(\hat{x}^{(c)}+\hbar D\left(\hat{x}^{(c)i}\right)),H_{2}(\hat{x}^{(c)}+\hbar D\left(\hat{x}^{(c)i}\right))]],
\end{equation}
where $\hat{O}^{(q)}$ represents a quantum operator and $H_{1}(\hat{x}^{(c)}+\hbar D(\hat{x}^{(c)i}))$
and $H_{2}(\hat{x}^{(c)}+\hbar D(\hat{x}^{(c)i}))$ are the corresponding
classical Hamiltonian operators.

In the Schr$\mathrm{\ddot{o}}$dinger representation, the time evolution
of a quantum state $|\psi\rangle$ can be expressed as:

\begin{equation}
i\frac{\partial}{\partial t}|\psi\rangle=\frac{1}{\hbar}[H_{1}(\hat{x}^{(c)}+\hbar D\left(\hat{x}^{(c)i}\right)),H_{2}(\hat{x}^{(c)}+\hbar D\left(\hat{x}^{(c)i}\right))]|\psi\rangle
\end{equation}
where $|\psi\rangle$ represents the quantum state.

These equations describe the time evolution of operators and states
within the framework of the operator formalism of classical Nambu
mechanics extended to include quantum operators and the Planck derivative.

By substituting the following expression for the Hamiltonians, 
\begin{equation}
H_{l}((\hat{x}^{(c)}+\hbar D\left(\hat{x}^{(c)i}\right))=H_{l}(\hat{x}^{(c)})+\frac{\partial H_{l}(\hat{x}^{(c)})}{\partial\hat{x}_{i}^{(c)}}\hbar D(\hat{x}_{i}^{(c)})+\dots,
\end{equation}
we obtain:
\begin{equation}
\frac{1}{\hbar}\begin{aligned}[][H_{1},H_{2}] & =+i\frac{1}{2}\epsilon_{ijk}\frac{\partial H_{1}(\hat{x}^{(c)})}{\partial\hat{x}_{i}^{(c)}}\frac{\partial H_{2}(\hat{x}^{(c)})}{\partial\hat{x}_{j}^{(c)}}\hat{\xi}_{k}-\frac{1}{2}\hbar\Delta_{Q},\end{aligned}
\end{equation}
where $\Delta_{Q}$ is defined as:

\begin{equation}
\Delta_{Q}\equiv[\frac{\partial H_{1}(\hat{x}^{(c)})}{\partial\hat{x}_{i}^{(c)}}D(\hat{x}_{i}^{(c)}),\frac{\partial H_{2}(\hat{x}^{(c)})}{\partial\hat{x}_{i'}^{(c)}}D(\hat{x}_{i'}^{(c)})]+\dots.
\end{equation}
Here, $\Delta_{Q}$ represents a quantum effect arising from the commutation
of the Planck derivative operator $D(\hat{x}^{(c)i})$ with the derivatives
of the classical Hamiltonians. The term $\frac{1}{2}\hbar\Delta_{Q}$
captures the quantum corrections to the classical bracket structure
and contributes to the overall evolution of the system.

Eventually, the following Schr$\mathrm{\ddot{o}}$dinger equation
is obtained:

\begin{equation}
i\frac{\partial}{\partial t}|\psi\rangle=\frac{[H_{1},H_{2}]}{\hbar}|\psi\rangle=i\frac{1}{2}\epsilon_{ijk}\frac{\partial H_{1}(\hat{x}_{1}^{(c)})}{\partial\hat{x}_{i}^{(c)}}\frac{\partial H_{2}(\hat{x}_{2}^{(c)})}{\partial\hat{x}_{j}^{(c)}}\hat{\xi}_{k}|\psi\rangle-\hbar\frac{1}{2}\Delta_{Q}|\psi\rangle+\dots.\label{eq:shrodinger}
\end{equation}

If we consider the position representation, we obtain:
\begin{equation}
\frac{\partial}{\partial t}\psi(x^{(c)},t)=\{H_{1},H_{2},\psi(x^{(c)},t)\}-\hbar\frac{1}{2}\int d^{3}x^{(c)}\langle x^{(c)}|\Delta_{Q}|x'^{(c)}\rangle\psi(x'^{(c)},t)+\dots.
\end{equation}

From this equation, the time evolution governed by Nambu brackets
is reproduced in the classical limit.

In Nambu quantum mechanics, the dual coordinate is introduced similarly:

\begin{equation}
\hat{\tilde{x}}^{(q)i}=\hat{x}^{(c)i}-\hbar D(\hat{x}^{(c)i}).
\end{equation}

The commutation relations for the dual coordinate are given by:

\begin{equation}
[\hat{\tilde{x}}^{(q)i},\hat{\tilde{x}}^{(q)j}]=-i\hbar\epsilon^{ijk}\hat{\xi}_{k},
\end{equation}

\begin{equation}
[\hat{\tilde{x}}^{(q)i},[\hat{\tilde{x}}^{(q)j},\hat{\tilde{x}}^{(q)k}]]=-i\hbar\epsilon^{ijk}.
\end{equation}
It should be noted that the dual coordinates also commute with the
original coordinates, 
\begin{equation}
[\hat{x}^{(q)i},\hat{\tilde{x}}^{(q)j}]=0.
\end{equation}

It is important to highlight the difference between Nambu quantum
mechanics and ordinary quantum mechanics. In Nambu quantum mechanics,
the operator $D$ is not explicitly defined in terms of the wave number
operator. To express $D$ explicitly, the dual operator must be introduced.
In this case, it can be written as:

\begin{equation}
\hbar D(\hat{x}^{(c)i})=\frac{\hat{x}^{(q)i}-\hat{\tilde{x}}^{(q)i}}{2}.
\end{equation}

The quantization of Nambu brackets was discussed in a two-step approach:
representing Nambu brackets in the operator formalism of classical
mechanics and introducing a quantum operator via the previously introduced
$D$.

The resulting commutation relation are:

\begin{equation}
[\hat{x}^{(q)i},[\hat{x}^{(q)j},\hat{x}^{(q)k}]]=i\hbar\epsilon^{ijk},
\end{equation}

\begin{equation}
[\hat{\tilde{x}}^{(q)i},[\hat{\tilde{x}}^{(q)j},\hat{\tilde{x}}^{(q)k}]]=-i\hbar\epsilon^{ijk},
\end{equation}

\begin{equation}
[\hat{x}^{(q)i},\hat{x}^{(q)j}]=i\hbar\epsilon^{ijk}\hat{\xi}_{k},
\end{equation}

\begin{equation}
[\hat{\tilde{x}}^{(q)i},\hat{\tilde{x}}^{(q)j}]=-i\hbar\epsilon^{ijk}\hat{\xi}_{k},
\end{equation}

\begin{equation}
[\hat{x}^{(q)i},\hat{\tilde{x}}^{(q)j}]=0.
\end{equation}
where $\hat{\xi}_{k}$ represents the wave number operator of $\hat{x}^{(c)i}$.

We note that a similar commutation relation appears in the context
of non-geometric string theory, specifically when considering the
T-duality of a closed string in a twisted torus with a B-field. In
this case, the commutation relation takes the form:

\begin{equation}
[\hat{x}^{i},\hat{x}^{j}]=\frac{il_{s}^{4}}{3\hbar}R^{ijk}\hat{p}_{k}
\end{equation}
where $l_{s}$ represents tne string length $l_{s}=\sqrt{\alpha'}$,
and $R^{ijk}$ is referred as the R-flux\cite{Szabo}.

In the theory of closed strings, the solution of the equation of motion
of a string is given by

\begin{equation}
X(\tau,\sigma)=X_{L}(\tau-\sigma)+X_{R}(\tau+\sigma)
\end{equation}
and its dual field is defined as:
\begin{equation}
\tilde{X}(\tau,\sigma)=X_{L}(\tau-\sigma)-X_{R}(\tau+\sigma).
\end{equation}

T-duality is achieved by interchanging the fields and dual fields:
\begin{equation}
X(\tau,\sigma)\longleftrightarrow\tilde{X}(\tau,\sigma).
\end{equation}

The zero-mode part of the fields can be expressed as:

\begin{equation}
X_{L}\left(\tau-\sigma\right)=x_{L}+\alpha'p_{L}(\tau-\sigma)+\dots,
\end{equation}

\begin{equation}
X_{R}(\tau-\sigma)=x_{R}+\alpha'p_{R}(\tau+\sigma)+\dots.
\end{equation}

Therefore, we have:

\begin{equation}
X(\tau,\sigma)=x+\alpha'\left(p\tau+\tilde{p}\sigma\right)+\dots,
\end{equation}

\begin{equation}
\tilde{X}(\tau,\sigma)=\tilde{x}+\alpha'\left(\tilde{p}\tau+p\sigma\right)+\dots,
\end{equation}
where 
\begin{equation}
x=x_{L}+x_{R},
\end{equation}

\begin{equation}
p=p_{L}+p_{R},
\end{equation}

\begin{equation}
\tilde{x}=x_{L}-x_{R},
\end{equation}
and

\begin{equation}
\tilde{p}=p_{L}-p_{R}.
\end{equation}
Here, $\tilde{p}$ represents the number of windings of the string
but corresponds to the momentum of the dual coordinate.

It should be noted that the commutation relation we discussed breaks
the Jacobi identity:

\begin{equation}
[\hat{x}^{(q)i},[\hat{x}^{(q)j},\hat{x}^{(q)k}]]+[\hat{x}^{(q)j},[\hat{x}^{(q)k},\hat{x}^{(q)i}]]+[\hat{x}^{(q)k},[\hat{x}^{(q)i},\hat{x}^{(q)j}]]=3i\hbar\epsilon^{ijk},
\end{equation}
and this breaking is solely due to the presence of the Planck constant.
This breaking of the Jacobi identity is analogous to the phenomenon
of Bianchi breaking by monopoles in gauge theory\cite{Jackiw}. In
the case of monopoles, it is discussed that the quantization of monopoles
should occur under a condition that avoids nonassociativity even if
the Jacobi law is not satisfied. Similarly, in our case, the breaking
of the Jacobi identity does not imply nonassociativity.

\section{Application}

In this section, we will explore some applications of the quantization
of Nambu brackets.

\subsection{Models that can be reduced to quantum mechanics}

In this section, we will discuss models that can be reduced to quantum
mechanics within the framework of Nambu brackets. Specifically, we
consider a scenario where one of the Hamiltonians, denoted as $x_{3}$,
leads to the Poisson bracket of ordinary classical mechanics. By choosing
$H_{1}$ as the harmonic oscillator, the Hamiltonians can be defined
as follows:

\begin{equation}
\begin{aligned}\hat{H}_{1} & =\frac{1}{2m}\left(\hat{x}_{2}^{(q)}\right)^{2}+\frac{k}{2}\left(\hat{x}_{1}^{(q)}\right)^{2},\\
\hat{H}_{2} & =\hat{x}_{3}^{(q)}.
\end{aligned}
\end{equation}

The Heisenberg equations of motion for this system are given by:

\begin{equation}
\begin{aligned}\dot{\hat{x}}^{(q)1} & =\frac{1}{i\hbar}[\hat{H}_{1},[\hat{H}_{2},\hat{x}^{(q)1}]]\\
 & =[\hat{H}_{1},\hat{\xi}_{2}]\\
 & =\frac{\hat{x}^{(q)2}}{m},
\end{aligned}
\end{equation}

\begin{equation}
\begin{aligned}\dot{\hat{x}}^{(q)2} & =\frac{1}{i\hbar}[\hat{H}_{1},[\hat{H}_{2},\hat{x}^{(q)2}]]\\
 & =-[\hat{H}_{1},\hat{\xi}_{1}]\\
 & =-k\hat{x}^{(q)1},
\end{aligned}
\end{equation}

\begin{equation}
\dot{x}^{(q)3}=\frac{1}{i\hbar}[\hat{H}_{1},[\hat{H}_{2},\hat{x}^{(q)1}]]=0.
\end{equation}

We observe that the equations of motion for $x^{(q)1}$ and $x^{(q)2}$
are identical to the equations of motion in quantum harmonic oscillators.
The time evolution of $x^{(q)3}$is constant, indicating that it remains
unchanged over time.

This is the same as the quantum theory of harmonic oscillators. Therefore,
in this particular case, the model reduces to the familiar quantum
theory of harmonic oscillators. This demonstrates the applicability
of the Nambu bracket formalism in capturing known quantum mechanical
systems and reproducing their dynamics.

\subsection{Lotka-Volterra Systems}

In this section, we consider the application of Nambu brackets to
describe Lotka-Volterra systems. It is known that the Lotka-Volterra
model can be described using Nambu brackets\cite{key-32}. The corresponding
Hamiltonians for this system are given by:

\begin{equation}
\begin{aligned}\hat{H}_{1} & =\hat{x}_{1}^{(q)}\hat{x}_{2}^{(q)}\hat{x}_{3}^{(q)},\\
\hat{H}_{2} & =\hat{x}_{1}^{(q)}+\hat{x}_{2}^{(q)}+\hat{x}_{3}^{(q)}.
\end{aligned}
\end{equation}

The Heisenberg equations of motion for this system are:

\begin{equation}
\begin{alignedat}{1}\dot{x}_{1}^{(q)} & =\frac{1}{i\hbar}[\hat{H}_{1},[\hat{H}_{2},\hat{x}_{1}^{(q)}]]\\
 & =\hat{x}_{1}^{(q)}\hat{x}_{3}^{(q)}-\hat{x}_{2}^{(q)}\hat{x}_{3}^{(q)},
\end{alignedat}
\end{equation}

\begin{equation}
\begin{alignedat}{1}\dot{x}_{2}^{(q)} & =\frac{1}{i\hbar}[\hat{H}_{1},[\hat{H}_{2},\hat{x}_{2}^{(q)}]]\\
 & =\hat{x}_{2}^{(q)}\hat{x}_{3}^{(q)}-\hat{x}_{1}^{(q)}\hat{x}_{2}^{(q)},
\end{alignedat}
\end{equation}

\begin{equation}
\begin{alignedat}{1}\dot{x}_{3}^{(q)} & =\frac{1}{i\hbar}[\hat{H}_{1},[\hat{H}_{2},\hat{x}_{3}^{(q)}]]\\
 & =\hat{x}_{1}^{(q)}\hat{x}_{2}^{(q)}-\hat{x}_{1}^{(q)}\hat{x}_{3}^{(q)}.
\end{alignedat}
\end{equation}

It is worth noting that these equations can also be written in the
following form: 
\begin{equation}
\begin{alignedat}{1}\dot{x}_{1}^{(q)}=\hat{x}_{3}^{(q)}\hat{x}_{1}^{(q)}-\hat{x}_{3}^{(q)}\hat{x}_{2}^{(q)}+i\hbar\left(-\hat{\xi}_{2}+\hat{\xi}_{1}\right)\end{alignedat}
,
\end{equation}

\begin{equation}
\begin{alignedat}{1}\dot{x}_{2}^{(q)}=\hat{x}_{1}^{(q)}\hat{x}_{2}^{(q)}-\hat{x}_{1}^{(q)}\hat{x}_{3}^{(q)}+i\hbar\left(-\hat{\xi}_{3}+\hat{\xi}_{2}\right)\end{alignedat}
,
\end{equation}
\begin{equation}
\begin{alignedat}{1}\dot{x}_{3}^{(q)}=\hat{x}_{1}^{(q)}\hat{x}_{2}^{(q)}-\hat{x}_{2}^{(q)}\hat{x}_{1}^{(q)}+i\hbar\left(-\hat{\xi}_{1}+\hat{\xi}_{3}\right)\end{alignedat}
,
\end{equation}
where we have used the commutation relation. 

It is interesting to note that the Lotka-Volterra equations can be
recovered from these equations of motion by neglecting the quantum
correction terms. The terms involving the wave number operators $\hat{\xi}_{i}$
can be treated as quantum corrections that arise from the quantization
of Nambu brackets.

By considering the classical limit where $\hbar$ approaches zero,
the quantum corrections vanish, and we are left with the classical
Lotka-Volterra equations. This demonstrates the correspondence between
the quantum description using Nambu brackets and the classical description
of the Lotka-Volterra system.

In summary, the quantization of Nambu brackets provides a framework
to describe Lotka-Volterra systems in a quantum mechanical setting.
The equations of motion obtained from Nambu brackets capture the dynamics
of the system, including both classical and quantum effects.

\section{Discussion}

In this paper, we present a new approach to the quantization of Nambu
brackets using operator formalism in classical mechanics. Our approach
introduces a Planck derivative (differential) $D$ that bridges classical
and quantum mechanics. By using $D$ to represent the Nambu brackets,
we derive a commutation relation that satisfies the quantization of
Nambu brackets. Interestingly, this commutation relation formally
coincides with the one manifested in the discussion of noncommutativity
in the context of double field theory.

The significance of this research is that it demonstrates how exploring
the correspondence between classical and quantum theory can contribute
to the quantization of the M2-brane and inspire further investigations
into the nature of quantum theory from a physical and fundamental
perspective. Reexamining the classical operator formalism and linking
it to quantum mechanics via the Planck derivative, as well as incorporating
the two-step process of converting to operator formalism and introducing
quantum corrections, offers a novel perspective on quantization.

Our research suggests several avenues for future exploration. One
such avenue is investigating the connection between our approach and
double field theory. Examining the potential implications of this
relationship could provide valuable insights into the fundamental
structure of M-theory and, consequently, the unification of quantum
mechanics and gravity.

Another promising direction is extending our approach beyond Nambu
brackets to encompass other complex structures. Further refinement
of the operator formalism may enhance our understanding of the quantization
processes in more intricate systems and facilitate a better comprehension
of the relationship between the classical and quantum realms.

\section*{Acknowledgments}

The author thanks Akio Sugamoto and Shiro Komata for their valuable
comments on this paper, and Yoshiki Matsuoka for reading it. 

\section*{Appendix A. Confirmation of commutation relation}

In this section, the calculation of the commutation relation by section
4 is detailed. For the sake of simplicity, we use symbols like the
following

\begin{equation}
Q=\hat{x}^{(q)1},\ P=\hat{x}^{(q)2},
\end{equation}

\begin{equation}
\tilde{Q}=\hat{x}^{(q)1},\tilde{P}=\hat{x}^{(q)2},
\end{equation}

\begin{equation}
Q_{c}=\hat{x}^{(c)1},\ P_{c}=\hat{p}^{(c)2},
\end{equation}

\begin{equation}
\xi_{q}=\hat{\xi}_{1},\xi_{p}=\hat{\xi}_{2}.
\end{equation}

Therefore, $Q$ ,P, $\tilde{Q},\tilde{P}$ are

\begin{equation}
Q=Q_{c}-\frac{1}{2}\xi_{p},
\end{equation}

\begin{equation}
P=P_{c}+\frac{1}{2}\xi_{q},
\end{equation}

\begin{equation}
\tilde{Q}=Q_{c}+\frac{1}{2}\xi_{p},
\end{equation}

\begin{equation}
\tilde{P}=P_{c}-\frac{1}{2}\xi_{q}.
\end{equation}

From the above, the commutation relation is calculated as follows

\begin{equation}
[Q,P]=[Q_{c}.\frac{1}{2}\xi_{q}]+[-\frac{1}{2}\xi_{p},P_{c}]=i,
\end{equation}

\begin{equation}
[\tilde{Q},\tilde{P}]=[Q_{c},-\frac{1}{2}\xi_{q}]+[\frac{1}{2}\xi_{p},P_{c}]=-i,
\end{equation}

\begin{equation}
[Q,\tilde{Q}]=[Q_{c}-\frac{1}{2}\xi_{p},Q_{c}+\frac{1}{2}\xi_{p}]=0,
\end{equation}

\begin{equation}
\begin{aligned}[][P,\tilde{Q}]= & [P_{c}+\frac{1}{2}\xi_{q},Q_{c}+\frac{1}{2}\xi_{p}]\\
= & [P_{c},\frac{1}{2}\xi_{p}]+[\frac{1}{2}\xi_{q},Q_{c}]=i\frac{1}{2}-i\frac{1}{2}=0,
\end{aligned}
\end{equation}

\begin{equation}
[P,\tilde{P}]=[P_{c}+\frac{1}{2}\xi_{q},P_{c}-\frac{1}{2}\xi_{q}]=0,
\end{equation}

\begin{equation}
\begin{aligned}[][Q,\tilde{P}] & =[Q_{c}-\frac{1}{2}\xi_{p},P_{c}-\frac{1}{2}\xi_{q}]\\
 & =[Q_{c},-\frac{1}{2}\xi_{q}]+[-\frac{1}{2}\xi_{p},P_{c}]=-\frac{i}{2}+\frac{i}{2}=0.
\end{aligned}
\end{equation}

\end{document}